# HILBERT MATRIX BASED CRYPTOSYSTEM USING A SESSION KEY

Penmetsa V Krishna Raja [#1,] A S N Chakravarthy [#2], Prof. P S Avadhani [#3]

[#1](Research Scholar, JNTUK, Kakinada, Andhra Pradesh, India, Email: vamsilovesindia@gmail.com)
[#2](Associate Professor, Department of CSE& IT, Sri Aditya Engineering College, Surampalem, Andhra Pradesh, India)
[#3](Professor, Dept. of CS & SE, Andhra University, Visakhapatnam, Andhra Pradesh, India)

**ABSTRACT**
**Cryptography protects users by providing functionality for the encryption of data and authentication of other users. This technology lets the receiver of an electronic message verify the sender, ensures that a message can be read only by the intended person, and assures the recipient that a message has not be altered in transit. Classical cryptanalysis involves an interesting combination of analytical reasoning, application of mathematical tools and pattern finding. The objectives of the proposed work are to propose a new cryptographic method based on the special matrix called the Hilbert matrix for authentication and confidentiality and to propose a model for confidentiality and authentication using shared key cryptosystems with the concept of digital enveloping using a session key. In the present work various algorithms are presented for encryption and authentication based on Hilbert matrix using a session key.**

## I. INTRODUCTION

Cryptography is a particularly interesting field because of the amount of work that is, by necessity, done in secret. The irony is that today, secrecy is not the key to the goodness of a cryptographic algorithm. Regardless of the mathematical theory behind an algorithm, the best algorithms are those that are well-known and well-documented because they are also well-tested and well-studied! In fact, time is the only true test of good cryptography; any cryptographic scheme that stays in use year after year is most likely a good one. The strength of cryptography lies in the choice (and management) of the keys; longer keys will resist attack better than shorter keys.

Session key is a secret used only once, that is, only for one session. This means that the secret key to be used by the sender and the receiver can be used only once. In other words, if a key has to shared by the sender and receiver then it should be used only when it is communicated and that too, only once. This can be done by a concept called digital enveloping. Digital enveloping is the concept of securely sending the session key. The session key will be encrypted with the public key of the receiver and sent. After receiving this, the receiver opens it with his private key and uses it. In the previous chapter, we have presented a symmetric encryption scheme based on Hilbert Matrix, in which the size of the Hilbert matrix is kept secret and known only to the sender and the receiver which means that n needs to be communicated totally through a secure channel. This is done using a session key. To transmit the session key with the concept of digital enveloping, RSA algorithm is used. In this paper two algorithms are presented for encryption and authentication based on Hilbert matrix using a session key.

## II. METHODOLOGY

The public key cryptosystems and the concepts of session key and the concept of the digital enveloping are studied. The public key systems include RSA, Diffie-Hellman Key exchange and NTRU. The underlying hard problems of number theory, abstract algebra and elliptic curves, namely, the factorization problem and the discrete logarithm problems are studied. While studying these, it is observed that Hilbert matrices have special properties which can be utilized for developing a symmetric cryptosystem. They are then analyzed and found suitable. A symmetric cryptosystem is designed, developed and implemented.

## III. CRYPTOSYSTEM BASED ON HILBERT MATRIX

The key generation, encryption and decryption of the Hilbert matrix cryptosystem are presented here. Here m is the block size and known to all.

3.1. Key Creation
Step 1: Choose a large prime number n and communicate securely between both the parties.

3.2. Encryption
Step 1: Let M be the plain text.
Step 2: Write the plaintext message as an m x 1 column matrix
Step 3: Choose a secret string K and encrypt with the public key of the receiver.
Step 4: Convert the result into a (n-m) x 1 column matrix  Step 5: Append this column matrix to make it an n x 1 column.
Step 6: Multiply the n x n Hilbert Matrix with this n x 1 column. The resultant is the cipher text.
Step 7: Transmit the cipher text to the receiver.

3.3 Decryption
    After receiving the cipher text by the receiver, he will decrypt it using the following algorithm.

Step 1: As the size n of the Hilbert Matrix is known, its inverse matrix is also known, which an n x n matrix is. Further, the cipher text is also known as an n x 1 column matrix.
Step 2: Multiply the inverse Hilbert matrix with the n x 1 cipher column matrix. The resultant contains the plain text and also the encrypted secret value of K
Step 3: Detach this and get the plain text. Observe that we need not decrypt the encrypted secret value as we are not interested in it. The algorithm is illustrated in figure 1.
In this diagram P is the plain text, K is a secret key, H is the Hilbert matrix and C is the cipher text

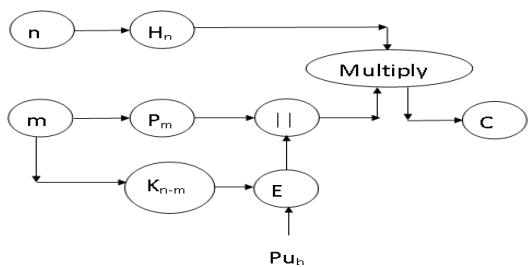

Fig. 1 Encryption using Hilbert matrix

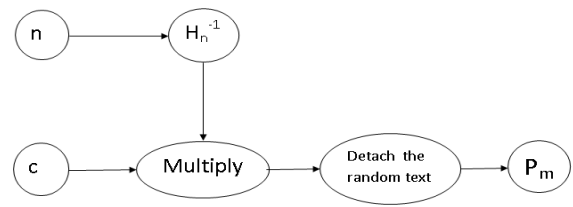

Fig. 2: Decryption using Hilbert matrix

The difficulty in this model is that the size n and also the secret string must be sent securely. Actually, it is sufficient if we send the size m of the plain text. It is really unnecessary to send the whole of the value K. We can use the concept of the public key cryptography to solve this by sending them as session keys. The basic idea in this is to send the sizes of the Hilbert matrix and the plain text strings as a session keys to the receiver.

## IV. ENCRYPTION ALGORITHM WITH A SESSION KEY CIPHER

Step 1: Select a large prime number n
Step 2: Encrypt n with the public key of the receiver. Call it n'
Step 3: Choose a number m less than n
Step 4: Encrypt m with the public key of the receiver. Call it m'
Step 5: Write the plaintext P as an m x 1 matrix
Step 6: Append any random string of n-m to P  and Call it T.
Step 7:  Compute Y = H T
Step 8:  Append 'n'of step 2 and m' of step 4 to  Y and
Call it C.
 Step 9: Transmit C to the other end.

## V. DECRYPTION ALGORITHM WITH A SESSION KEY

After receiving the cipher text C, it may be decrypted by the receiver using the following algorithm.
Step 1: First separate Y, m' and n' from C.
Step 2: Decrypt n' and m' using the private key of the receiver to get the values of m' and n'.
Step 3: Compute $H^{-1}Y$ to get T.
Step 4: Now the size m of the garbage is known   and hence the garbage is also known.
Step 5: Remove the garbage to get plain text P.

It can be seen from the algorithm that the sender and the receiver need not agree upon the size of the Hilbert matrix (and hence the Hilbert matrix itself) because it is transported as a session key. Thus size can be varied for each session. Further the size of the plain text is also sent as an encrypted value, using the concept of digital enveloping.

The introduction of secret random garbage of the size n-m serves two purposes.
1) As the size of the secret key is unknown, the size n of the Hilbert matrix is kept unknown and
2) It can also be used for authentication of the message which is presented in the following algorithm.
In the algorithm mentioned in section IV, only n is used as a session key but the random string is kept secret although we have not made use of it. This can also be used as a session key for authentication. This is presented in algorithm in algorithm mentioned in section VI.

## VI. ALGORITHM FOR AUTHENTICATION BASED ON HILBERT MATRIX

Step 1: Select a large prime number n
Step 2: Encrypt n with the public key of the receiver. Call it n'
Step 3: Choose a number m less than n
Step 4: Encrypt m with the public key of the receiver. Call it m'
Step 5: Choose a random string K of size n-m
Step 6: Encrypt K with the public key of the receiver. Call it K' and send it to the receiver.
Step 7: Write the plaintext P as an m x 1 matrix
Step 8: Append K to P and call it T.
Step 9: Compute Y = H T
Step 10: Append n' of step 2 and m' of step 4 to Y and Call it C.
Step 11: Transmit C to the other end. Upon receiving C, the receiver decrypts it as follows.
Step 12: First separate Y, m' and n' from C.
Step 13: Decrypt n' and m' and using the private key of the receiver to get the values of m' and n'.
Step 14: Compute $H^{-1}Y$ to get T.
Step 15: Now the size n-m is known and hence the K is also known.

Step 16: Encrypt K with the public key of the receiver and compare it with K', which is already received.
Step 17: If the comparison is successful the message is authenticated otherwise it is rejected.
Step 18: If K is removed from T the plaintext P is obtained.

Here the garbage value is being used as some sort of hash value for authentication. The algorithm is implemented in Java and the results are satisfactory.

## VII. SHARED KEY CRYPTOSYSTEM USING HILBERT MATRIX

The concept of shared key has been used various cryptosystems like RSA [17]. Basically any shared key cryptosystem is applicable when there are multiple participants. That is, if a message has be securely transmitted by k members then shared key concepts are being used. For example, two or more people can have a bank locker and if it has to be opened only when all of them agree to open it. It is simple application of shared key. In this chapter, a shared key algorithm with k participants using Hilbert matrix is presented.

7.1 Sum Protocol
The encryption methodology for a point to point communication based on the Hilbert Matrix has already been discussed in the previous chapter. In the encryption scheme, the size n of the Hilbert matrix is kept secret and communicated either through a secure channel or shared key.
In the proposed shared key algorithm, n is a shared key sent by all the k participants. This n will be sum of all the secret value of all the participants. Now the challenge lies in designing an algorithm in such a way that each participant i will have a secret value $n_i$ (known only to him) such that $n_1+n_2+n_3+…+n_k = n$. we have to design a protocol such that ni is known only to the i[th] participant but not to anybody else and n is known to all. The protocol works as follows:
Assume that there are k participants. Each participant i will have a secret number $n_i$.

Step 1: Each participant select a secret number $n_i$ of his choice.
Step 2: Participant one will select another secret number c of his choice.
Step 3: Participant one will $n_1+c$ and sends it to participant two.
Step 4: Participant two will add $n_2$ to the number received from participant one and sends it to participant three.
Step 5: Participant three adds $n_3$ to the received number and sends it to participant four.
Step 6: Continue this process till the last participant k and participant will add $n_k$ and sends it back to participant one.

Step 7: Now participant one subtracts the secret value C from value he received from participant k and broadcast it to all the participants which will be the sum n.

7.2 Shared Key encryption using Hilbert matrix
The share key encryption and decryption using Hilbert matrix cryptosystem are presented here. Here it is assumed that there are k participants and the message will be sent only if all the k participants agree. The encryption and decryption algorithms for shared key are presented here.
Step 1: Each participant i will chose a secret number $n_i$ of his choice and kept secret.
Step 2: Using sum protocol discussed above, find the sum n of all secret values $n_i$   Hence $n = n_1 + n_2 + .... + n_k$. Step 3: Use algorithms in [16] for encryption and decryption.

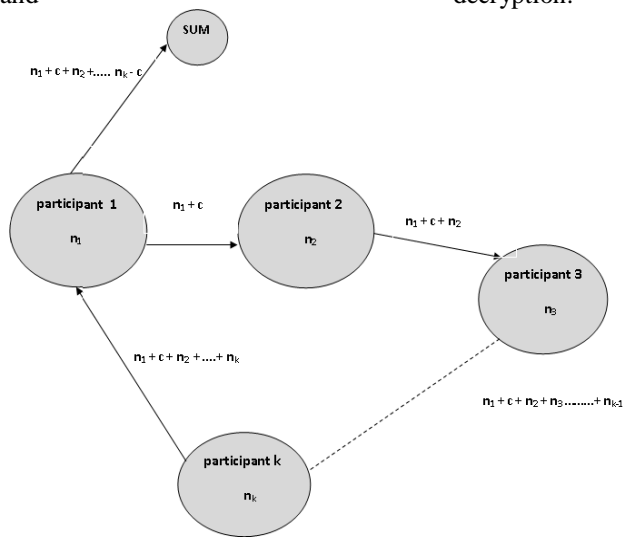

**Fig. 3 Sum Protocol**

## VIII.  CONCLUSIONS & FUTURE DIRECTIONS

In the present work a new cryptosystem based on Hilbert Matrices, which is an improvement [8] of has been proposed. The idea behind choosing the Hilbert matrices is that they are unstable and have integer inverses [1]. These properties make them interesting and they are listed below.
1. Whatever may be the order, Hilbert matrices are invertible.
2. The inverse of a Hilbert matrix will have all its entries integers.
3. If the order is known then the inverse can be easily found and it is very difficult to find the inverse if the order is unknown.
4. Direct computation of the inverse of the Hilbert matrix leads to more round off errors due to its unstable nature.

All these properties proved to be an advantage for the designing a cryptosystem based on Hilbert matrix. As the size of the Hilbert matrix is kept secret (known only to sender and receiver) because of the instability and the point 4 above, it is difficult and practically impossible for anyone to retrieve the message without knowing n.

Further, as the inverse has closed form solution with all its entries being integers, the round of errors can be avoided. As n needs to be kept secret and the plain text size is supposed to be public, we need to have another string to be augmented to the plain text .This is also kept secret and this secret string K is used as a session key and also can be used for authentication.

The share key algorithm presented here can be extended for authentication in a shared environment further in the place of secret random garbage value, any key or equivalent can be transmitted securely. These can be used in applications like copyright protection and digital water marking because of its simplicity and ease of use. This can be extended to other models of security like the group key communication.

## AUTHORS

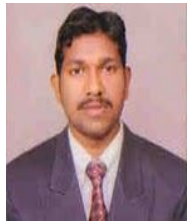

[1]Penmetsa V Krishna Raja received his M.Tech (CST) from A.U, Visakhapatnam, Andhra Pradesh, India. He is a research scholar under the supervision of Prof.P.S.Avadhani. His research areas include Network Security, Cryptography, Intrusion Detection, Neural networks.

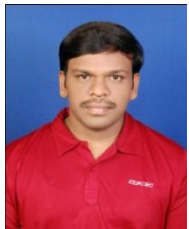

[2]A.S.N Chakravarthy received his M.Tech (CSE) from JNTU, Anantapur , Andhra Pradesh, India. Presently he is working as an Associate Professor in Dept. Of Computer Science and Engineering in Sri Aditya Engineering College, SuramPalem, AP, India. He is a research scholar under the supervision of Prof.P.S.Avadhani His research areas include Network Security, Cryptography, Intrusion Detection, Neural networks, Digital Forensics and Cyber Security.

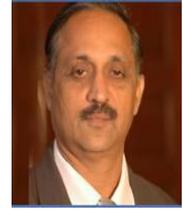

[3]Prof. P.S.Avadhani did his Masters Degree and PhD from IIT, Kanpur. He is presently working as Professor in Dept. of Computer Science and Systems Engineering in Andhra University college of Engg., in Visakhapatnam. He has more than 50 papers published in various National / International journals and conferences. His research areas include Cryptography, Data Security, Algorithms, and Computer Graphics, Digital Forensics and Cyber Security.